\documentstyle{mn}

\newif\ifAMStwofonts

\ifoldfss
  \ifCUPmtlplainloaded \else
    \NewTextAlphabet{textbfit} {cmbxti10} {}
    \NewTextAlphabet{textbfss} {cmssbx10} {}
    \NewMathAlphabet{mathbfit} {cmbxti10} {} % for math mode
    \NewMathAlphabet{mathbfss} {cmssbx10} {} %  "   "    "
  \fi
  \ifAMStwofonts
    \ifCUPmtlplainloaded \else
      \NewSymbolFont{upmath} {eurm10}
      \NewSymbolFont{AMSa} {msam10}
      \NewMathSymbol{\upi}     {0}{upmath}{19}
      \NewMathSymbol{\umu}     {0}{upmath}{16}
      \NewMathSymbol{\upartial}{0}{upmath}{40}
      \NewMathSymbol{\leqslant}{3}{AMSa}{36}
      \NewMathSymbol{\geqslant}{3}{AMSa}{3E}

      \let\leq=\leqslant 
       
    \fi
  \fi
\fi % End of OFSS

\ifnfssone
  \newmathalphabet{\mathit}
  \addtoversion{normal}{\mathit}{cmr}{m}{it}
  \addtoversion{bold}{\mathit}{cmr}{bx}{it}
  \newmathalphabet{\mathbfit} % math mode version of \textbfit{..}
  \addtoversion{normal}{\mathbfit}{cmr}{bx}{it}
  \addtoversion{bold}{\mathbfit}{cmr}{bx}{it}
  \newmathalphabet{\mathbfss} % math mode version of \textbfss{..}
  \addtoversion{normal}{\mathbfss}{cmss}{bx}{n}
  \addtoversion{bold}{\mathbfss}{cmss}{bx}{n}
  \ifAMStwofonts
    \ifCUPmtlplainloaded \else
      \UseAMStwoboldmath
      \makeatletter
      \new@mathgroup\upmath@group
      \define@mathgroup\mv@normal\upmath@group{eur}{m}{n}
      \define@mathgroup\mv@bold\upmath@group{eur}{b}{n}
      \edef\UPM{\hexnumber\upmath@group}
      \new@mathgroup\amsa@group
      \define@mathgroup\mv@normal\amsa@group{msa}{m}{n}
      \define@mathgroup\mv@bold\amsa@group{msa}{m}{n}
      \edef\AMSa{\hexnumber\amsa@group}
      \makeatother
      \mathchardef\upi="0\UPM19
      \mathchardef\umu="0\UPM16
      \mathchardef\upartial="0\UPM40
      \mathchardef\leqslant="3\AMSa36
      \mathchardef\geqslant="3\AMSa3E

      \let\leq=\leqslant 

    \fi
  \fi
\fi % End of NFSS release 1

\ifnfsstwo
  \DeclareMathAlphabet{\mathbfit}{OT1}{cmr}{bx}{it}
  \SetMathAlphabet\mathbfit{bold}{OT1}{cmr}{bx}{it}
  \DeclareMathAlphabet{\mathbfss}{OT1}{cmss}{bx}{n}
  \SetMathAlphabet\mathbfss{bold}{OT1}{cmss}{bx}{n}
  \ifAMStwofonts
    \ifCUPmtlplainloaded \else
      \DeclareSymbolFont{UPM}{U}{eur}{m}{n}
      \SetSymbolFont{UPM}{bold}{U}{eur}{b}{n}
      \DeclareSymbolFont{AMSa}{U}{msa}{m}{n}
      \DeclareMathSymbol{\upi}{0}{UPM}{"19}
      \DeclareMathSymbol{\umu}{0}{UPM}{"16}
      \DeclareMathSymbol{\upartial}{0}{UPM}{"40}
      \DeclareMathSymbol{\leqslant}{3}{AMSa}{"36}
      \DeclareMathSymbol{\geqslant}{3}{AMSa}{"3E}

      \let\leq=\leqslant 

    \fi
  \fi
\fi % End of NFSS release 2

\ifCUPmtlplainloaded \else
  \ifAMStwofonts \else % If no AMS fonts
    \def\upi{\pi}
    \def\umu{\mu}
    \def\upartial{\partial}
  \fi
\fi

\title[CASE: RR Lyrae stars from the globular cluster $\omega$ Centauri 
as standard candles]
{Cluster AgeS Experiment (CASE): ~RR Lyrae stars from \linebreak
the globular cluster $\omega$ Centauri as standard candles}
\author[A. Olech et. al.]
{A.~Olech,$^{1}$~  J.~Kaluzny,$^{1}$~
 I.B.~Thompson$^2$ and A. Schwarzenberg-Czerny$^{1,3}$\\
  $1$Nicolaus Copernicus Astronomical Center,
ul. Bartycka 18, 00-716 Warsaw, Poland (olech,jka,alex@camk.edu.pl)\\
  $^2$Carnegie Institution of Washington, 813 Santa Barbara Street,
Pasadena,CA 91101, USA (ian@ociw.edu)\\
  $^3$Adam Mickiewicz University Observatory, ul. Sloneczna 36,
60-286 Poznan, Poland\\
}

\date{Accepted 2003 .................., Received 2003 March 1}

\pagerange{\pageref{firstpage}--\pageref{lastpage}}
\pubyear{2003}

\begin{document}

\maketitle

\label{firstpage}

\begin{abstract}

New photometry of RRab and RRc stars in $\omega$ Cen is used to
calibrate their absolute magnitudes $M_V$ as a function of: a)
metallicity; and b) the Fourier parameters of light curves in the $V$
band. The zero point of both calibrations relies on the distance
modulus to the cluster derived earlier by the CASE project based on
observations of the detached eclipsing binary OGLE~GC17. For RRab
variables we obtained a relation of $M_{\rm V}= (0.26\pm 0.08)[{\rm
Fe/H}]+(0.91\pm 0.13)$. A dereddened distance modulus to the LMC based
on that formula is $\mu_{0}=18.56\pm 0.14$ mag. The second calibration
of $M_{\rm V}$, which is based on Fourier coefficients of decomposed
light curves, results in the LMC distance of $\mu_{0}=18.51\pm 0.07$ mag.

\end{abstract}


\section{Introduction}

The Cluster AgeS Experiment (CASE) is a long-term project whose main
goal is to determine the distances and ages of nearby globular clusters
(Thompson et al. 2001). The primary distance indicators used by the CASE
group are detached eclipsing binaries. The search for these relatively
rare objects is performed using 1-meter class telescopes. Usually for
each target cluster, several hundred CCD frames are collected over one or
more observing seasons. Suitable candidates are then observed with
larger telescopes (eg. Thompson et al. 2001). A secondary benefit of the
project's survey phase is the detection of large samples of variables of
various types including RR~Lyr and SX~Phe stars.

The first globular cluster for which the CASE project determined a
distance was $\omega$ Centauri. An analysis of photometric and
spectroscopic data for an eclipsing binary OGLE~GC17 yielded an apparent
distance modulus equal to $(m-M)_V=14.09\pm0.04$ mag (Kaluzny et al.
2002).

Time series photometry of $\omega$ Cen obtained by the CASE project in
the 1999 and 2000 seasons resulted in light curves of almost 400
variables from the cluster field (Kaluzny et al. 2003). The most
numerous group were the RR Lyr stars. The new precise and well-sampled
$BV$ light curves of these stars can be supplemented with information
about their metallicities (Rey et al. 2000). In this paper we use these
data to calibrate different formulae which permit the calculation of
absolute magnitudes of RR~Lyr variables.

\section{$M_V - {\rm [Fe/H]}$ relation}

It has been known for many years that the absolute magnitude of RR Lyr
stars $M_V$ is a function of metallicity [Fe/H] (Sandage 1981a,b). This
function is often assumed to be a simple linear relation of the form:

\begin{equation}
M_V = \alpha {\rm [Fe/H]} + \beta
\end{equation}

There is a lack of general consensus on the exact values of the $\alpha$
and $\beta$ parameters. The most extreme values of the slope of the
relation (1) were given by Sandage (1981a,b) with $\alpha=0.35$ and Fusi
Pecci et al. (1996) with $\alpha=0.13\pm0.07$. The estimates of the zero
point of the relation vary from around 1.1 mag  (Gould \& Popowski 1998)
to 0.8 mag (Gratton et al. 1997, Caloi et al. 1997).

It is worth pointing out that some theoretical models predict that
a simple linear relationship of form (1) may not exist at all. This is
caused by the fact that even variables of the same metallicity have
different luminosities depending on the  direction of their
evolution across the instability strip (Lee 1991). More complex
relations between $M_V$ of RR Lyr stars and their metallicities were
documented, for example, by Caputo (1997).

There is also observational evidence for a non-linear dependence of
$M_V$ on [Fe/H]. For example, Caputo et al. (2000), analyzing RR Lyr
variables from Galactic globular clusters, obtained
$\alpha=0.17\pm0.04$ for ${\rm [Fe/H]}<-1.5$ and $\alpha=0.27\pm0.06$
for ${\rm [Fe/H]}>-1.5$. The systematic errors caused by using the
simple linear relation (1) may be minimized by excluding highly evolved
RR Lyraes from the analyzed samples. In particular, the formula should be
used with caution when analyzing variables belonging to globular
clusters showing "blue" horizontal branches.

Recent investigation of the $M_V - {\rm [Fe/H]}$ relation in $\omega$
Cen was done by Rey et al. (2000). They used [Fe/H] metallicities
derived from the  $hk$ index of the Ca$by$ photometric system. Their
intensity averaged magnitudes of RR Lyr stars were taken from
photographic photometry from Butler et al. (1978) and the CCD photometry
of Kaluzny et al. (1997).

The new $V$-band light curves of RR Lyr variables in $\omega$ Cen
obtained by the CASE (Kaluzny et al. 2003) have about three times as
many observed points as the photometry of Kaluzny et al (1997). This
permits a more precise estimation of mean $<V>$ magnitudes and allows
the elimination of objects with unstable light curves from the calibration
sample. Adopting an apparent distance modulus of the cluster as
determined by (Kaluzny et al. 2002) we can obtain a new $M_V - {\rm
[Fe/H]}$ relation for RR Lyr stars in $\omega$ Cen.

The sample used for the calibration includes 122 stars with stable light
curves of good quality and metallicities determined by Rey et al.
(2000). We decided to remove the RRab variable V52 from the further
analysis. Its high luminosity suggests that it is a foreground star.
Additionally, van Leeuwen et al. (2000) gives only a 45\% membership
probability for this variable, based on a proper motion study.

The $M_V - {\rm [Fe/H]}$ dependence for remaining 121 variables is shown
in the upper panel of Fig. 1. The middle panel shows the same relation
but for a sub-sample consisting of 67 RR$ab$ variables. The errors in
[Fe/H] were taken directly from Table 5 of Rey et al. (2000). For
objects with unknown error of [Fe/H] we assumed it to be equal to 0.2
dex as suggested by Rey et al. (2000).

The intensity averaged magnitudes $<V>$ obtained from light curves
containing 500-800 points have internal errors smaller than 0.001 mag.
We estimate that the external error of the zero point of the photometry
is about 0.02 mag.

\begin{figure}
 \vspace{13.6cm}
 \caption{The relation between $M_{\rm V}$ and  [Fe/H] for:
a) the whole sample; b) sub-sample of RR$ab$ variables; and c)
for unevolved RR$ab$ variables. The solid lines are best linear fits.}
\includegraphics{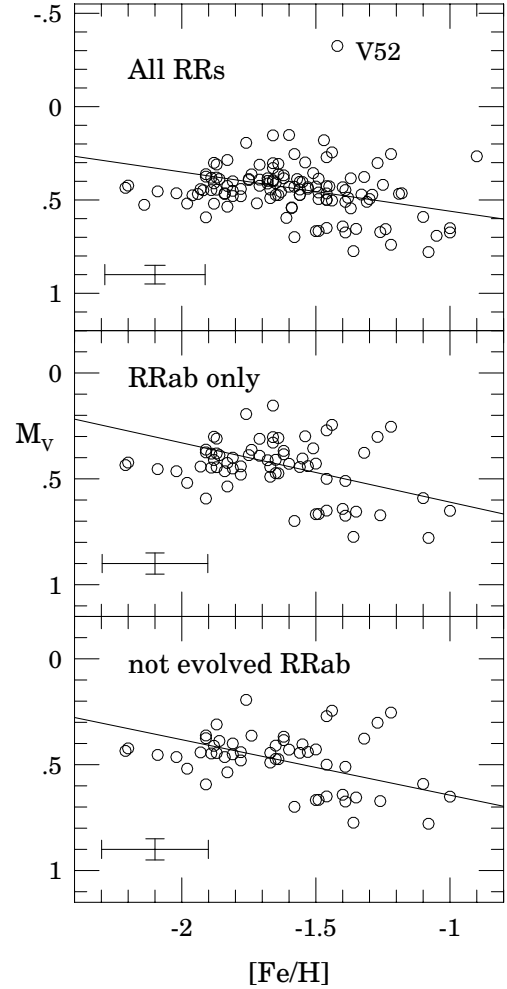}
\end{figure}

Transformation from $<V>$ to $M_V$ is performed using the distance
modulus whose error is 0.04 mag. Thus, combining this value with 0.02
mag error resulting from computing the mean magnitudes $<V>$, we
assumed that individual absolute magnitudes of RR Lyr stars from our
sample have uncertainties of 0.05 mag.

Knowing the errors we were able to fit straight lines to the graphs
shown in Fig. 1. For the sample of 121 RR Lyr stars we obtain the
following relation:

\begin{equation}
M_V = (0.21\pm0.05)\cdot {\rm [Fe/H]} + (0.77\pm0.08)
\end{equation}

\noindent and for 67 RR$ab$ variables is in the form:

\begin{equation}
M_V = (0.28\pm0.07)\cdot {\rm [Fe/H]} + (0.89\pm0.12)\\
\end{equation}

The values of $\alpha$ derived above may be biased by the presence of
extremely evolved objects in our sample. The evolutionary models of Lee
(1990) suggest that the RR Lyrae stars in clusters having a very
blue HB and with a metallicity in the range of $-2.0 < {\rm [Fe/H]} < -1.6$
are highly evolved stars. They have significantly brighter magnitudes and
longer periods than those near the zero age horizontal branch (ZAHB).

Rey et al. (2000) investigated this problem in $\omega$ Cen. Analyzing
the period-amplitude relations for different ranges of metallicities
they discovered that a significant sample of evolved RR Lyr stars exists
only for variables with $-1.9 \leq {\rm [Fe/H]} < -1.5$. Looking at
their Fig. 9b one can clearly see that these evolved objects are also
characterized by periods longer than 0.7 days.

Thus we decided to exclude from our sample all RR$ab$ variables having
metallicities in a range of $-1.9 \leq {\rm [Fe/H]} < -1.5$ and pulsation
periods longer than 0.7 days. Our final sample consists of 53 RR$ab$
variables shown in the lower panel of Fig. 1. 
The resulting relation for
$M_V$ is as follows:

\begin{equation}
M_V = (0.26\pm0.08)\cdot {\rm [Fe/H]} + (0.91\pm0.13)
\end{equation}

The slope of this relation falls inside the 0.13-0.35 range presented in
the literature. Recently, Chaboyer (1999) in his review of globular
cluster distance determinations adopted $\alpha=0.23\pm0.04$ arguing
that 1-$\sigma$ range of this value encompasses the majority of recent
determinations for $\alpha$.

Our results do not differ significantly from those presented by Rey et
al. (2000), due to the good agreement of the zero points of the
CASE photometry, as well as the CCD and photographic photometry used by Rey et. al. (2000). The mean difference in $V$ magnitude between both data sets
consisted of 53 RR$ab$ stars is $<V_{CASE} - V_{Rey}> = 0.007 \pm 0.010$ 
mag.

Having established the relation (4), we can use it for determining the distance modulus to the Large Magellanic Cloud - $\mu_{LMC}$. We used the data of Walker (1992), who summarized the CCD photometry of 182 RR Lyr variables
belonging to seven LMC globular clusters. We decided to exclude from our
analysis variables from the cluster NGC 1841 because it is most probably
located about 0.3 mag closer than the main body of the LMC. The mean
value of the reddening free magnitude $<V_0>$ of 160 RR Lyr stars from
the remaining six clusters is $18.98\pm0.03$ mag. The mean metallicity
of this sample is ${\rm [Fe/H]}=-1.9\pm0.2$.

Knowing the observed magnitude of RR Lyr stars from the LMC and
computing their absolute magnitude using relation (4) we determined
that the distance modulus to the LMC is equal to
$\mu_{LMC}=18.57\pm0.20$ mag. This result supports the "long" distance
scale.

Most recently, Clementini et al. (2003) presented new photometry and
spectroscopy for more than a hundred RR Lyrae stars in two fields
located close to the bar of the LMC. The dereddened apparent average
luminosity of the variables from their sample was $<V_0> = 19.064 \pm
0.064$ mag at the mean metal abundance of ${\rm [Fe/H]} = -1.48\pm0.03$.
Their metallicity was tied to the Harris (1996) scale which differs by
about 0.06 dex from the Zinn \& West (1984) scale used by Rey et al.
(2000). Thus the mean metallicity of their sample is ${\rm [Fe/H]} =
-1.54\pm0.03$ on the Zinn \& West (1984) scale.

Application of equation (4) for their data gives the distance modulus to
the LMC equal to $\mu_{LMC}=18.55\pm0.19$ mag.

Combining the results of photometry obtained by both Walker (1992) and
Clementini et al. (2003) we find that our 
$M_V$-${\rm [Fe/H]}$ relation gives $\mu_{LMC}=18.56\pm0.14$ mag.

\section{Absolute magnitude of RR Lyr stars as a function of the Fourier
parameters}\label{s3}

\subsection{RR$ab$ stars}\label{s31}

\subsubsection{Methods}\label{s311}

On the one hand, nonlinear pulsation models of RR Lyr stars suggested
that the luminosity of these stars may be uniquely related to
the pulsation period and the shape of their light curves (e.g.
Simon \& Clement 1993). This strongly suggests that there is a close
correlation between the Fourier coefficients of the light curves and their corresponding periods and luminosities. Such correlations could potentially improve the distances derived for these stars.

On the other hand serious attempts to find such empirical
relations pioneered by Kov\'acs \& Jurcsik (1996, hereafter KJ96)
and most recently updated by Kov\'acs \& Walker (2001) - hereafter
KW01 - bore mixed success. The empirical procedure is essentially
reduced to adopting a linear combination of the period and $M$
selected Fourier parameters ${\cal F}$ for predicting the RR
Lyr absolute magnitude:
\begin{equation}
M_V(P{\cal F})=c_0+c_P\log{P}+\sum^{M}_{m=1} c_m {\cal
F}_m\label{e:1}
\end{equation}
The coefficients $c_m$ and cluster distance moduli $d_k$ are
fitted to minimize scatter $\chi^2$ in the apparent average
magnitudes $V_{kn}$ summed over $N_k$ RR Lyr stars in $K$
clusters:
\begin{equation}
\chi^2=\sum_{k=1}^K\sum_{n=1}^{N_k}[V_{kn}-d_k-M_V(P{\cal
F})]^2\label{e:1a}
\end{equation}
Problems are indicated by an apparent lack of convergence from the
whole procedure to a unique solution. Different data sets yield
substantially different relations, both in respect to the optimum
parameter set and in the numerical values of the involved
coefficients. For RRab stars KJ96 obtained:
\begin{equation}
M_V = 1.221 - 1.396{P} - 0.477 A_1 + 0.103
\varphi_{31}\label{e:2}
\end{equation}
while KW01 using 383 RR$ab$ stars from 20 globular clusters
recommend either of three:
\begin{eqnarray}
M_V &=& -1.820 \log{P} - 0.805 A_1 + c \label{e:3}\\ M_V &=&
-1.876 \log{P} - 1.158 A_1 + 0.821 A_3 + c \label{e:4}\\ M_V &=&
-1.963 \log{P} - 1.124 A_1 + 0.830 A_3 + \nonumber \\ ~& ~&
~~~~~~~~~~~~+0.011 \phi_{51} + c\label{e:5}
\end{eqnarray}

Note that the sine phase convention is used here unless otherwise
stated. A worrying inconsistency is shown, for example, by the values of
$c_P$ and $c_{A1}$ in Eq. (\ref{e:2}) and (\ref{e:3}).

Several different causes might be responsible for these problems:
\begin{enumerate}
\item The original problem connected to the sensitivity of high
order Fourier coefficients to poor data melted away with abundance of
the high quality observations (e.g. Kaluzny et al. 1997, KW01 and
the present work).

\item The more persistent problem is connected to the apparent inability
of the procedure to account for the hidden diversity of properties
related to metallicity, evolutionary status and possibly misidentified
pulsation mode, all capable of producing several different trends in the
data. This problem is acknowledged by rejecting outlying stars from the
rest by the $D_m$ distance criterion (see e.g. Kov\'acs \& Kanbur 1998).
The risk is that the omitted stars hide much information on the nature
of the physics involved. 

\item One reason for widely different values of the coefficients
obtained from the different data sets could be the strong correlation
between the Fourier coefficients. Such fits could be sensitive to a group of
coefficients while changing loads between individual members of the group
matters little.

\item Finally, some evidence discussed further in this paper
indicates that magnitudes of RRab stars might also depend on 
a factor, so far unaccounted for, independent of the period and shape of the
light curve. We recommend future trials with colors sensitive to
temperature and/or metallicity.
\end{enumerate}

In order to clarify the situation and possibly to identify the cause of
these problems we undertook the same kind of analysis from scratch and
for an entirely new set of data. Our present sample consists of 76
RR$ab$ variables observed in $\omega$ Cen by Kaluzny at el. (1997) and
Kaluzny et al. (2003). The photometry in these two papers is of very
good quality and shows no systematic differences in the zero point (see
Kaluzny et al. 2003 for details). Following recommendations of Kov\'acs
\& Kanbur (1998) for analysis we used only $N_{\omega Cen}=56$ stars
with $D_m<3$. A clear advantage of our data is very consistent photometry
obtained within the same project and using the same telescope. The vast
majority of stars was observed at least 500 times, hence we obtained very
precise and reliable values of the Fourier coefficients. 

\begin{figure*}
 \vspace{18.4cm}
 \caption{The amplitudes $A_j$, amplitudes ratios $R_{j1}$ and
phase combinations $\phi_{j1}$ in the function of the period $P$ for all
RR$ab$ variables from our sample.}
\includegraphics{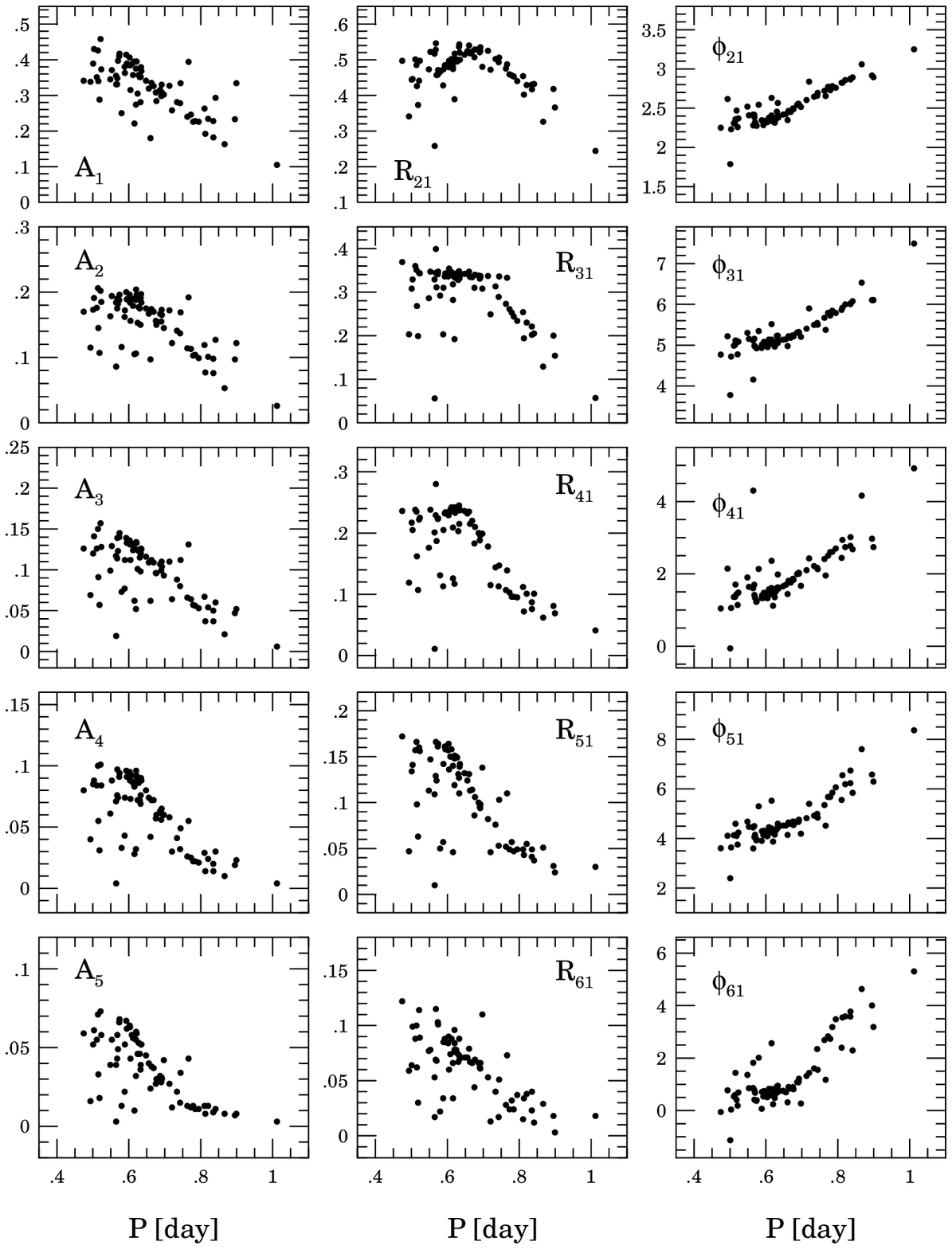}
\end{figure*}

%%\clearpage

\begin{figure*}
 \vspace{22cm}
 \caption{The $M_V - <V>$ relations for the different formulae connecting
the absolute magnitudes of RR$ab$ stars with their light curve parameters.
Solid lines in all panels have a slope of unity. Filled circles correspond
to $M_V$ values computed from our formulae and open circles to the formulae
of KW01. The $M_V$ values obtained from KW01 equations are shifted by 0.4 mag
for clarity.}
\includegraphics{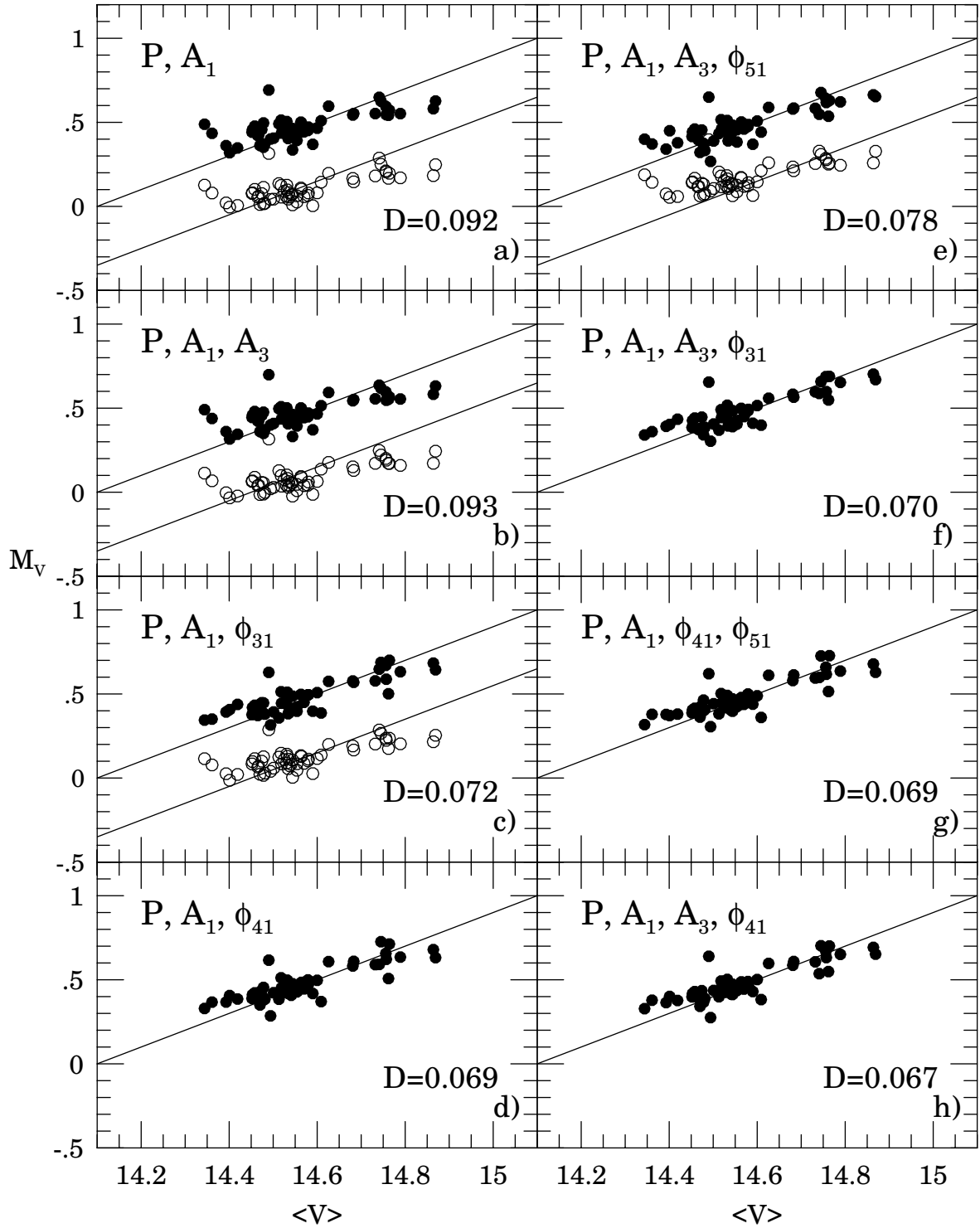}
\end{figure*}

%%\clearpage

\begin{table*}
 \centering
 \begin{minipage}{200mm}
  \caption{Basic parameters of RR$ab$ in $\omega$ Cen based on the photometry
of Kaluzny et al. (1997) and Kaluzny et al. (2003).}
  \begin{tabular}{rccccccccccc}
\hline
Star & $P$ [days] & $<V>$ & $m_0$ & $A_1$ & $A_2$ & $A_3$ & $\phi_{31}$
& $\phi_{41}$ & $\phi_{51}$ & $D_m$ & $n_{obs}$ \\
\hline
V4 & 0.627320 & 14.453 & 14.499 & 0.357 & 0.177 & 0.123 & 5.074 & 1.504 & 4.323 & 1.30 & 591\\
V5 & 0.515274 & 14.745 & 14.803 & 0.340 & 0.145 & 0.091 & 5.114 & 1.703 & 4.596 & 1.49 & 593\\
V7 & 0.713037 & 14.590 & 14.633 & 0.327 & 0.172 & 0.110 & 5.405 & 2.101 & 4.820 & 2.07 & 211\\
V8 & 0.521329 & 14.683 & 14.756 & 0.458 & 0.202 & 0.157 & 4.775 & 1.139 & 3.755 & 1.70 & 757\\
V9 & 0.523480 & 14.756 & 14.804 & 0.373 & 0.185 & 0.128 & 5.085 & 1.483 & 4.241 & 1.78 & 591\\
V13 & 0.669039 & 14.465 & 14.500 & 0.326 & 0.170 & 0.109 & 5.183 & 1.757 & 4.592 & 1.48 & 592\\
V18 & 0.621689 & 14.570 & 14.619 & 0.395 & 0.196 & 0.133 & 5.072 & 1.487 & 4.327 & 1.62 & 756\\
V20 & 0.615559 & 14.579 & 14.631 & 0.395 & 0.190 & 0.126 & 5.136 & 1.608 & 4.408 & 2.83 & 715\\
V23 & 0.510870 & 14.869 & 14.911 & 0.351 & 0.176 & 0.126 & 4.989 & 1.360 & 4.131 & 1.21 & 713\\
V25 & 0.588364 & 14.478 & 14.520 & 0.380 & 0.162 & 0.077 & 4.933 & 1.327 & 3.905 & 1.37 & 589\\
V27 & 0.615680 & 14.757 & 14.770 & 0.221 & 0.105 & 0.062 & 5.516 & 2.360 & 5.525 & 1.41 & 589\\
V32 & 0.620373 & 14.530 & 14.586 & 0.395 & 0.204 & 0.134 & 5.064 & 1.488 & 4.319 & 1.36 & 720\\
V33 & 0.602324 & 14.544 & 14.601 & 0.406 & 0.196 & 0.136 & 5.050 & 1.429 & 4.236 & 1.98 & 591\\
V34 & 0.733967 & 14.481 & 14.508 & 0.281 & 0.141 & 0.088 & 5.497 & 2.214 & 4.920 & 2.05 & 592\\
V38 & 0.779061 & 14.478 & 14.493 & 0.226 & 0.103 & 0.057 & 5.728 & 2.603 & 5.680 & 1.75 & 746\\
V40 & 0.634072 & 14.519 & 14.572 & 0.379 & 0.197 & 0.126 & 5.131 & 1.621 & 4.500 & 1.73 & 567\\
V41 & 0.662942 & 14.537 & 14.572 & 0.335 & 0.174 & 0.112 & 5.220 & 1.809 & 4.646 & 1.33 & 742\\
V44 & 0.567545 & 14.732 & 14.772 & 0.331 & 0.175 & 0.114 & 5.158 & 1.699 & 4.502 & 1.16 & 582\\
V45 & 0.589116 & 14.531 & 14.574 & 0.363 & 0.172 & 0.112 & 5.015 & 1.401 & 4.295 & 1.49 & 591\\
V46 & 0.686971 & 14.501 & 14.537 & 0.315 & 0.167 & 0.107 & 5.309 & 1.993 & 4.693 & 2.14 & 591\\
V49 & 0.604627 & 14.609 & 14.630 & 0.315 & 0.156 & 0.112 & 4.954 & 1.326 & 4.107 & 1.30 & 587\\
V51 & 0.574152 & 14.564 & 14.616 & 0.417 & 0.196 & 0.145 & 4.924 & 1.238 & 3.969 & 1.87 & 590\\
V54 & 0.772915 & 14.419 & 14.437 & 0.246 & 0.113 & 0.064 & 5.792 & 2.502 & 5.677 & 2.36 & 591\\
V56 & 0.568023 & 14.762 & 14.781 & 0.348 & 0.190 & 0.139 & 4.985 & 1.416 & 4.056 & 2.13 & 592\\
V59 & 0.518506 & 14.741 & 14.760 & 0.288 & 0.107 & 0.057 & 5.105 & 1.437 & 4.107 & 1.49 & 569\\
V62 & 0.619770 & 14.474 & 14.520 & 0.375 & 0.188 & 0.125 & 5.080 & 1.492 & 4.313 & 1.09 & 588\\
V67 & 0.564451 & 14.681 & 14.726 & 0.355 & 0.184 & 0.117 & 5.116 & 1.600 & 4.448 & 0.79 & 594\\
V74 & 0.503209 & 14.626 & 14.694 & 0.430 & 0.191 & 0.141 & 4.722 & 1.055 & 3.638 & 1.16 & 738\\
V79 & 0.608276 & 14.600 & 14.656 & 0.388 & 0.189 & 0.132 & 5.139 & 1.569 & 4.454 & 1.70 & 280\\
V85 & 0.742758 & 14.470 & 14.496 & 0.278 & 0.137 & 0.080 & 5.543 & 2.167 & 4.995 & 2.72 & 233\\
V86 & 0.647844 & 14.541 & 14.582 & 0.341 & 0.175 & 0.116 & 5.135 & 1.656 & 4.521 & 1.62 & 729\\
V90 & 0.603404 & 14.525 & 14.575 & 0.384 & 0.184 & 0.131 & 4.993 & 1.329 & 4.089 & 2.60 & 713\\
V96 & 0.624527 & 14.344 & 14.374 & 0.305 & 0.153 & 0.101 & 4.965 & 1.354 & 4.157 & 1.39 & 588\\
V97 & 0.691898 & 14.534 & 14.563 & 0.308 & 0.165 & 0.104 & 5.314 & 2.025 & 4.760 & 2.26 & 729\\
V100 & 0.552745 & 14.789 & 14.839 & 0.371 & 0.194 & 0.129 & 5.155 & 1.640 & 4.475 & 0.91 & 735\\
V102 & 0.691396 & 14.554 & 14.592 & 0.330 & 0.173 & 0.110 & 5.319 & 2.000 & 4.695 & 2.14 & 724\\
V106 & 0.569903 & 14.518 & 14.574 & 0.397 & 0.182 & 0.123 & 4.965 & 1.334 & 4.151 & 1.41 & 674\\
V107 & 0.514102 & 14.864 & 14.927 & 0.426 & 0.206 & 0.150 & 5.006 & 1.363 & 4.132 & 1.60 & 665\\
V108 & 0.594458 & 14.532 & 14.586 & 0.389 & 0.189 & 0.133 & 5.081 & 1.478 & 4.313 & 2.04 & 739\\
V111 & 0.762905 & 14.393 & 14.410 & 0.240 & 0.114 & 0.066 & 5.671 & 2.409 & 5.351 & 2.41 & 710\\
V112 & 0.474359 & 14.490 & 14.531 & 0.341 & 0.170 & 0.126 & 4.770 & 1.042 & 3.607 & 1.71 & 748\\
V113 & 0.573375 & 14.563 & 14.621 & 0.410 & 0.188 & 0.140 & 4.935 & 1.224 & 3.927 & 1.45 & 720\\
V114 & 0.675307 & 14.467 & 14.499 & 0.308 & 0.156 & 0.096 & 5.244 & 1.864 & 4.686 & 0.97 & 745\\
V115 & 0.630474 & 14.535 & 14.578 & 0.351 & 0.175 & 0.115 & 5.230 & 1.607 & 4.381 & 1.74 & 588\\
V118 & 0.611618 & 14.458 & 14.500 & 0.357 & 0.179 & 0.124 & 5.033 & 1.435 & 4.243 & 1.17 & 568\\
V120 & 0.548537 & 14.764 & 14.803 & 0.345 & 0.163 & 0.099 & 5.302 & 1.900 & 4.682 & 1.40 & 575\\
V122 & 0.634929 & 14.554 & 14.601 & 0.368 & 0.184 & 0.125 & 5.102 & 1.592 & 4.474 & 1.42 & 577\\
V125 & 0.592888 & 14.580 & 14.653 & 0.414 & 0.200 & 0.139 & 5.016 & 1.336 & 4.140 & 2.22 & 261\\
V132 & 0.655656 & 14.452 & 14.487 & 0.319 & 0.167 & 0.109 & 5.147 & 1.696 & 4.524 & 1.65 & 563\\
V135 & 0.632579 & 14.513 & 14.540 & 0.281 & 0.150 & 0.098 & 5.052 & 1.623 & 4.429 & 1.97 & 562\\
V139 & 0.676871 & 14.361 & 14.387 & 0.284 & 0.150 & 0.096 & 5.216 & 1.834 & 4.542 & 1.75 & 589\\
V141 & 0.697363 & 14.494 & 14.519 & 0.302 & 0.145 & 0.093 & 5.205 & 1.668 & 4.193 & 1.04 & 569\\
V144 & 0.835320 & 14.401 & 14.411 & 0.182 & 0.076 & 0.037 & 6.006 & 3.014 & 6.742 & 2.79 & 586\\
V268 & 0.812922 & 14.544 & 14.555 & 0.192 & 0.077 & 0.037 & 5.917 & 2.939 & 6.552 & 2.75 & 586\\
\hline
\end{tabular}
\end{minipage}
\end{table*}

To minimize errors in the coefficients and their correlation we converted our
projection into orthogonal trigonometric polynomials
(Schwarzenberg-Czerny 1996, Schwarzenberg-Czerny \& Kaluzny 1996). As
all stars come from the same cluster we do not suffer from the distance
indeterminacy and differential interstellar reddening because there are
no systematic differences in $E(B-V)$ in the field of $\omega$ Cen
(Schlegel et al. 1998). Additionally, because of the significant spread
in metallicities within $\omega$ Cen, our sample is excellent for
studying metallicity related intra-cluster differences unaffected by
distance differences.

Figure 2 shows the amplitudes $A_j$, amplitude ratios $R_{j1}$ and
phase combinations $\phi_{j1}$ as the function of the period $P$
for all RR$ab$ variables from our sample. Comparing them with the
same relations for the KW01 sample (see their Fig. 1) we conclude that
our relations are significantly tighter and correlate very well
with the period.

The interesting thing we noted is the clear change of the slope in the
$\phi_{51}$ - $P$ and $\phi_{61}$ - $P$ relations occurring at
$P\approx0.75$ days. Arranging the RR$ab$ stars from $\omega$ Cen
according to the increasing period one can see that it is connected with
a disappearance of the "bump", clearly seen at phase around 0.7 in the light
curves of variables with periods shorter than 0.75 days.

The basic properties of our sample of RR$ab$ stars from $\omega$
Cen are collected in Table 1.

\subsubsection{Results for $\omega$ Cen}\label{s312}

We performed several fits involving $P$ and up to 3 different Fourier
phase and amplitude coefficients drawn from the first 5 harmonics. To
remove the constant term indeterminacy, we adopted as fixed the distance
modulus $\mu_{\omega Cen} = 14.09\pm0.04$ mag of Kaluzny et al. (2002).
The results are listed in Table 2. We explicitly list the fitted formula
and its standard deviation $\cal D$. To test the numerical self
consistency of our fits we use each formula to recalculate $\mu_{\omega
Cen}$ and its standard deviation $\sigma$. These are also listed in
Table 2. Its hardly surprising that ${\cal D}/\sigma \approx
\sqrt{N_{\omega Cen}}$. We shall refer to different fitted formulae by
their consecutive numbers in the Table 2, from (F1) to (F11).

Inspection of Table 2 shows that no formulae involving only amplitudes
(e.g. F1 \& F5) are satisfactory. This is in marked contradiction to
expectations from Eqs. (\ref{e:3}) and (\ref{e:4}). It is particularly
surprising as the latter formula performed best on the data from KW01. We
observe a strong correlation effect of type (iii) in Sect. \ref{s311}.
In F1 \& F5 only the amplitude coefficients vary while the rest remains
remarkably stable, and the quality of the fit $(\cal D)$ is unaffected.
In F5, contributions from $A_1$ and $A_3$ simply cancel out to the value
in F1 for $A_1$ alone. Most other fits involving {\em both} amplitude
and phase appear satisfactory. It seems that the set of parameters
originally selected by KJ96 performs quite well in our case, except that
our coefficients in F3 are completely different from those in Eq.
(\ref{e:2}). Formally, our best fit is F6. However, the improvement of
$\chi^2$ for F6 compared to that for F3 yields a Fisher-Snedecor test
value of $F(1,52)=7.85$, only marginally significant at the level of 99\%.
At the face of these values one could say that F2, F3, F6--F10 yield
fits of similar quality and inclusion of more parameters are hardly
justifiable. It looks like the whole procedure is incapable of yielding
better accuracy than these latter formulae.

In order to reveal any remaining trend in the residuals from the
fit in each panel of Fig. 3 we plot for all stars their absolute
magnitude $M_V$ predicted by a given formula against average
apparent magnitude $<V>$. Dots correspond to the formulae from
Table 2 and circles to Eqs. (\ref{e:3}--\ref{e:5}) from KW01, the
latter shifted by a constant. The lines mark ideal relations $M_V
= <V>-\mu_{\omega Cen}$. Inspection of Fig. 3 reveals that
compared to Eqs. (\ref{e:3}--\ref{e:5}) our formulae involving
phase perform particularly well for brighter stars. The common
feature in all panels is the markedly lower inclination of the trend
in the points than the ideal relation. This is worrisome as
apparently the fitted formulae are unable to reproduce the
observed span of magnitudes. This effect is also present in
similar attempts to predict magnitudes of cepheids using shapes of
their light curves (Ligeza \& Schwarzenberg-Czerny 2000). In our
opinion the presence of such an effect might indicate that some
hidden factor (different from the period and shape of the light
curve) significantly influences the average magnitudes of RRab stars
(see point iv in Sect. \ref{s311}).

\subsubsection{Application to LMC}\label{s312}

\begin{table*}
 \centering
 \begin{minipage}{200mm}
  \caption{Formulae for $M_V$ of RR$ab$ stars with their ${\cal D}$ parameters and
the resulting distance moduli to $\omega$ Cen with its standard deviations}
  \begin{tabular}{rlccc}
\hline \hline No & Equation & ${\cal D}$ & $\mu_{\omega Cen}$ &
$\sigma$ \\ \hline \hline 1& $M_V = 0.272 - 1.868 \log P - 0.542
A_1$ & 0.0920 & 14.0900 & 0.0123 \\ \hline 2& $M_V = -0.970 -
3.055 \log P + 0.734 A_1 + 0.333 \phi_{41}$ & 0.0692 & 14.0903 &
0.0092
\\ 3& $M_V = -2.858 - 3.016 \log P + 0.529 A_1 + 0.488 \phi_{31}$ &
0.0722 & 14.0917 & 0.0096 \\ 4& $M_V = -0.911 - 2.601 \log P +
0.327 A_1 + 0.163 \phi_{51}$ & 0.0785 & 14.0892 & 0.0104 \\ 5&
$M_V = 0.286 - 1.872 \log P - 0.717 A_1 + 0.410 A_3$ & 0.0928 &
14.0895 & 0.0123
\\ \hline 6& $M_V = -1.000 - 3.155 \log P + 0.095 A_1 + 1.698 A_3 +
0.356 \phi_{41}$ & 0.0673 & 14.0907 & 0.0088 \\ 7& $M_V = -0.822 -
3.114 \log P + 0.756 A_1 + 0.453 \phi_{41} - 0.082 \phi_{51}$ &
0.0690 & 14.0899 & 0.0090 \\ 8&$M_V = -3.135 - 3.162 \log P -
0.211 A_1 + 2.002 A_3 + 0.542 \phi_{31}$ & 0.0696 & 14.0902 &
0.0091 \\ 9&$M_V = -1.214 - 3.068 \log P + 0.729 A_1 + 0.058
\phi_{31} + 0.298 \phi_{41}$ & 0.0699 & 14.0918 & 0.0092 \\
10&$M_V = -2.873 - 3.017 \log P + 0.528 A_1 + 0.493 \phi_{31} -
0.002 \phi_{31}$ & 0.0730 & 14.0899 & 0.0096 \\ 11&$M_V = -0.946 -
2.666 \log P - 0.219 A_1 + 1.413 A_3 + 0.174 \phi_{51}$ & 0.0776 &
14.0916 & 0.0102 \\ \hline \hline
\end{tabular}
\end{minipage}
\end{table*}

To verify the usefulness of the formula F6 we decided to determine
the distance modulus of the LMC. To do this we required high
quality photometry of RR Lyr variables from the LMC. The first
source we verified was the photometry of RR Lyr variables in seven
globular clusters placed in the LMC (Walker 1992). Unfortunately,
these light curves contained only about 30 points, which is
insufficient to obtain valuable information about high order
Fourier coefficients.

The photometry of 68~000 variable stars in the LMC was performed
by the OGLE team and recently published by \.Zebru\'n et al. (2001).
Unfortunately, due to the fact that OGLE searches for 
microlensing phenomena, the vast majority of their data is collected in
the $I$-band. Additionally, the exposure times are too short to produce
very high quality data on RR Lyr variables. The photometry described by
\.Zebru\'n et al. (2001) was collected in 21 fields and for most of them
the number of $V$-band frames was around 30-40. Only four fields,
namely SC2, SC3, SC4 and SC5, were observed over 50 times in $V$ and thus
the light curves of RR Lyr stars located in these fields have
photometry suitable for our tests.

The total number of RR$ab$ stars in the above-mentioned four fields of
the LMC was 1428. Only 13 of them have light curves characterized by the
parameter $D_m<3$ and we used our F6 equation only for these. Such a
large number of rejected stars cannot remain without discussion because
it could suggest that our formulae are incapable of predicting the
behavior of LMC RRab stars. The OGLE observations of RR Lyr variables
are only a by-product of a project designed for other purposes, thus the
number of $V$ filter observations, their exposure length and
distribution are not optimal for measuring of such fine effects as the
high harmonics and amplitudes of the light curves of RR$ab$ stars.

Leaving aside the problem of a large number of rejected stars, we
proceeded to apply the remaining 13 stars to determine the
LMC distance modulus. As we are concerned with the zero point
only, the problems with the $M_V$ modeling and intrinsic scatter of
stars pose less danger. The resulting $M_V - <V>$ relation for
these 13 variables is shown in Fig. 4. Again the solid line has a
slope of unity. Averaging the differences between the observed and
absolute magnitudes of the stars from this sample we obtain the
value of the apparent distance modulus to the LMC as equal to
$18.83\pm0.04$ mag, where the error is a standard deviation of the
mean of 13 estimates.

\begin{figure}
 \vspace{7cm}
 \caption{The $M_V - <V>$ relation for the 13 OGLE RR$ab$ variables
from the LMC. The solid line has a slope of unity.}
\includegraphics{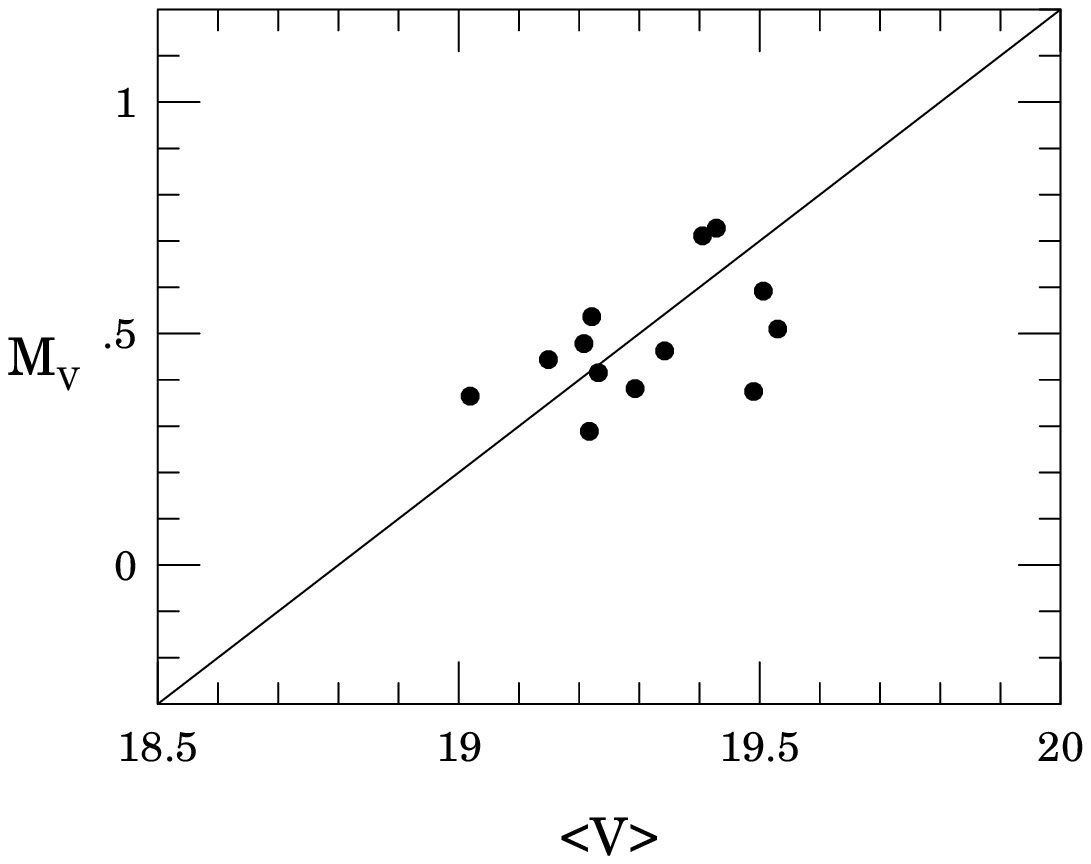}
\end{figure}

In order to obtain the true distance modulus to the main body of
the LMC, knowledge about interstellar reddening is needed.
Due to the significant angular size of the LMC in the sky, the
reddening varies depending on the position. However, the OGLE fields
which we use are near the center of the LMC and adopting
the average value of $E(B-V)$ taken from recent estimates of
reddening should not introduce large systematic errors. As was
shown, for example, by Alves et al. (2002), Fitzpatrick et al.
(2002), Dutra et al. (2001) and Groenewegen \& Salaris (2001) the
interstellar reddening toward the LMC varies from 0.086 to 0.12
mag. Thus we simply assume that $E(B-V)$ is equal to $0.10\pm0.02$
mag and our estimate of the true distance modulus to the LMC is
$\mu_{LMC}=18.51$ mag. Taking into account the standard deviation
of the mean value of our 13 estimates, the error in the absorption
in $V$-band introduced by the error in $E(B-V)$ and the error in
calibrating the zero point from the relation F6 of our final
estimate of the distance to the LMC is
$\mu_{LMC}=18.51\pm0.07_r\pm0.04_s$ mag.

\begin{figure*}
 \vspace{22cm}
 \caption{The amplitudes $A_j$, amplitudes ratios $R_{j1}$ and
phase combinations $\phi_{j1}$ in the function of the period $P$ for all
RR$c$ variables from our sample.}
\includegraphics{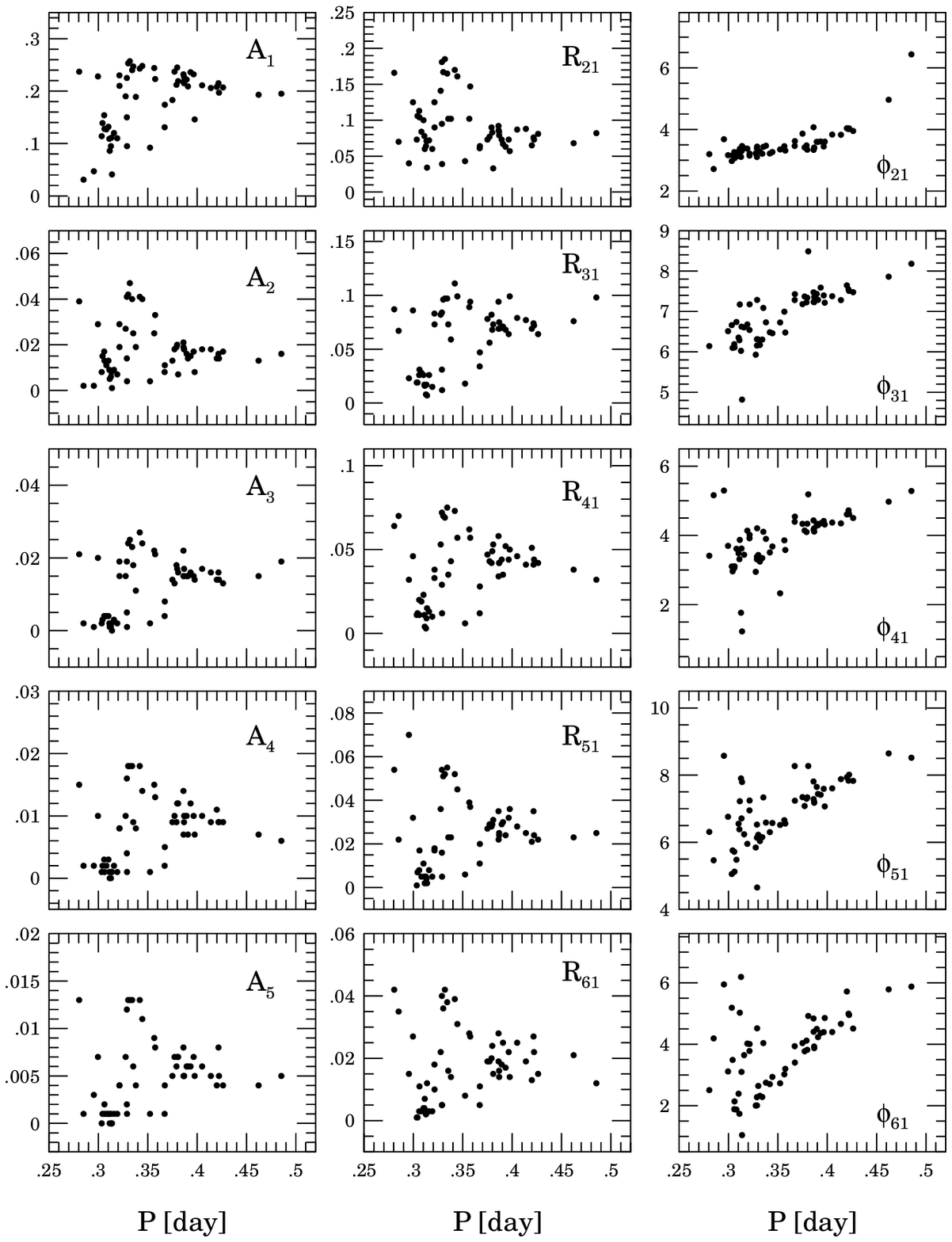}
\end{figure*}

\subsection{RR$c$ stars: near sinusoidal pulsators}\label{s32}
\subsubsection{Is the luminosity-shape correlation real for RR$c$?}\label{s321}

In fact theoretical investigation of correlations among physical
parameters of stars and their pulsation light curve for RRc stars
started before those for RRab. Investigations by Simon \& Teays
(1982) and Simon (1989) relying on a large number of hydrodynamic
pulsation models indicated that $L$ correlates with $\phi_{31}$
and predominantly with $P$. Simon \& Clement (1993) came out with
the following formula for RRc luminosity:
\begin{equation}
\log L = 1.04 \log P - 0.058 \phi^*_{31} + 2.41\label{e:11}
\end{equation}
where $\phi^*_{31}$ indicates cosine phase convention, differing
from our sine convention by $\pi$. The form of this formula
resembles Eq. (\ref{e:2}) except for a linear transformation of
units.
\begin{figure}
 \vspace{15.5cm}
 \caption{The $M_V - <V>$ relations for the different formulae connecting
the absolute magnitudes of RR$c$ stars with their light curve parameters.
Solid lines in all panels have a slope of unity.}
\includegraphics{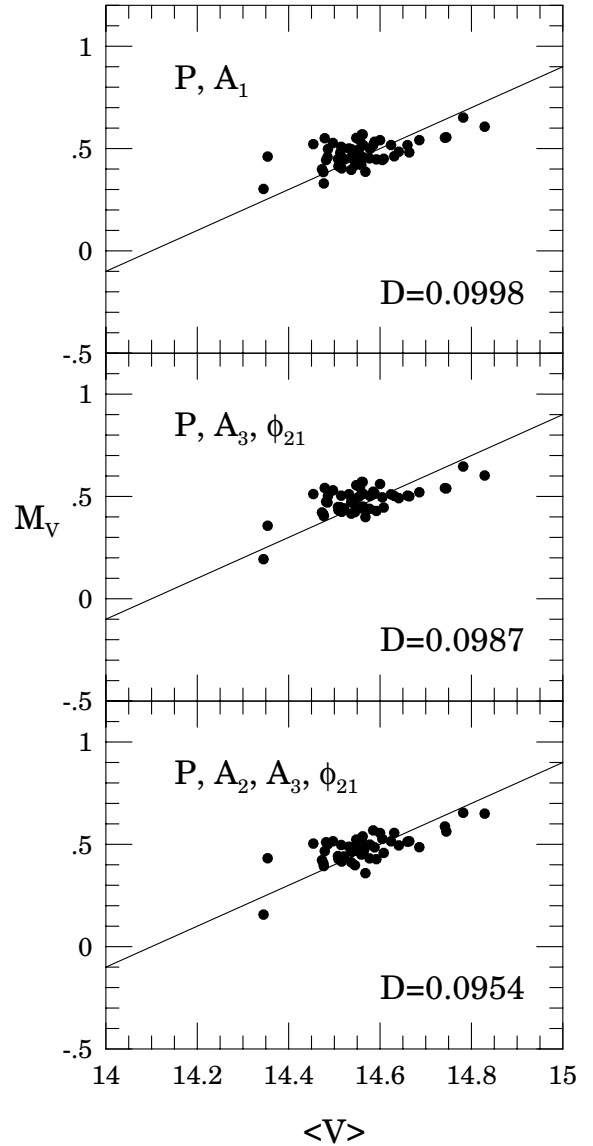}
\end{figure}

Our sample contains 54 RR$c$ stars from $\omega$ Cen observed by Kaluzny
et al. (1997, 2003) within the OGLE and CASE projects. For this sample
we selected only stars free of any complications, such as period
changes, multiple periods, or low-amplitude noisy light curves. For the
RRc stars we followed the same procedure as for the RRab. First we
determined their Fourier parameters and then proceeded to fit
formulae akin to Eq. (\ref{e:1}) by minimizing $\chi^2$. On the one hand,
the quality of our photometry for RRc is as good as ever. On the other
hand, the small amplitude and nearly sinusoidal light curve of RRc stars
produced increased errors for high harmonics, so that above the 6-th
harmonic they exceeded 10\% of the value. The periods, magnitudes and
Fourier parameters of the RR$c$ stars from our sample are listed in
Table \ref{t:3}. The amplitudes $A_j$, amplitudes ratios $R_{j1}$ and
phase combinations $\phi_{j1}$ in the function of the period $P$ for all
RR$c$ variables from our sample are shown in Fig. 5.

These data were fitted with formulae of the type of Eq. (\ref{e:2}),
using $P$ and up to 3 Fourier coefficients derived for up to the 4-th
harmonic. The results were a bit disappointing in that all types of
formulae yielded rather large standard deviations and of similar order
${\cal D}\approx 0.1$ mag. In Table \ref{t:4} we list selection of the
best formulae for each length category. It must be kept in mind, that
according to the $F$ test, none of these formulae performed
significantly better than the rest, at 0.95 level. Note that the phase
term present in formulae F2 and F3 does not vary by more than 0.1 mag.
For all these reasons we recommend the use of the simplest F1 formula:

\begin{equation}
M_V = -0.259 - 1.338 \log P - 0.726 A_1\label{e:12}
\end{equation}

The corresponding $M_V - <V>$ relations shown in Fig. 6 reveal
that the predicted $M_V$ vary across fewer than half of the
observed range. It is rather disturbing to note that all stars
except for the 3 most outlying ones form a broad horizontal clump
consistent with no correlation of $M_V$ with $V$. It is hard to
imagine a more vivid demonstration that the observed luminosity
and shape of the RRc light curve do not follow as tight a relation
as the theoretical one (see point iv in Sect. \ref{s311}).

\begin{table*}
 \centering
 \begin{minipage}{200mm}
  \caption{Basic parameters of RR$c$ in $\omega$ Cen based on the photometry
of Kaluzny et al. (1997) and Kaluzny et al. (2003).\label{t:3}}
  \begin{tabular}{rcccccccccc}
\hline
Star & $P$ [days] & $<V>$ & $m_0$ & $A_1$ & $A_2$ & $A_3$ & $\phi_{21}$
& $\phi_{31}$ & $\phi_{41}$ & $n_{obs}$ \\
\hline
V10 & 0.374976 & 14.482 & 14.491 & 0.183 & 0.013 & 0.014 & 3.863 & 7.181 & 4.334 & 590\\
V12 & 0.386769 & 14.521 & 14.532 & 0.215 & 0.018 & 0.015 & 3.411 & 7.229 & 4.203 & 590\\
V14 & 0.377114 & 14.513 & 14.528 & 0.237 & 0.018 & 0.013 & 3.414 & 7.373 & 4.142 & 592\\
V16 & 0.330202 & 14.562 & 14.577 & 0.253 & 0.042 & 0.024 & 3.307 & 6.306 & 3.434 & 743\\
V19 & 0.299551 & 14.829 & 14.841 & 0.228 & 0.029 & 0.020 & 3.164 & 6.512 & 3.698 & 752\\
V21 & 0.380812 & 14.354 & 14.366 & 0.219 & 0.007 & 0.016 & 1.050 & 8.491 & 5.186 & 589\\
V24 & 0.462278 & 14.477 & 14.487 & 0.193 & 0.013 & 0.015 & 4.963 & 7.864 & 4.973 & 730\\
V35 & 0.386841 & 14.562 & 14.574 & 0.227 & 0.018 & 0.017 & 3.399 & 7.358 & 4.110 & 726\\
V36 & 0.379846 & 14.545 & 14.558 & 0.245 & 0.020 & 0.017 & 3.341 & 7.343 & 4.337 & 228\\
V39 & 0.393374 & 14.565 & 14.579 & 0.236 & 0.015 & 0.016 & 3.607 & 7.593 & 4.349 & 735\\
V47 & 0.485303 & 14.345 & 14.354 & 0.195 & 0.016 & 0.019 & 0.156 & 8.184 & 5.283 & 592\\
V50 & 0.386172 & 14.631 & 14.644 & 0.232 & 0.021 & 0.022 & 4.074 & 7.477 & 4.433 & 591\\
V64 & 0.344497 & 14.556 & 14.569 & 0.248 & 0.040 & 0.024 & 3.266 & 6.461 & 3.680 & 591\\
V70 & 0.390687 & 14.556 & 14.568 & 0.209 & 0.014 & 0.015 & 3.596 & 7.290 & 4.294 & 708\\
V71 & 0.357544 & 14.532 & 14.545 & 0.223 & 0.033 & 0.021 & 3.313 & 6.479 & 3.579 & 724\\
V75 & 0.422174 & 14.476 & 14.484 & 0.197 & 0.014 & 0.014 & 4.036 & 7.511 & 4.617 & 718\\
V76 & 0.337962 & 14.515 & 14.524 & 0.189 & 0.019 & 0.011 & 3.471 & 6.730 & 3.899 & 713\\
V77 & 0.426294 & 14.568 & 14.579 & 0.207 & 0.017 & 0.013 & 3.951 & 7.480 & 4.502 & 707\\
V81 & 0.389392 & 14.608 & 14.620 & 0.222 & 0.016 & 0.015 & 3.598 & 7.439 & 4.336 & 754\\
V83 & 0.356612 & 14.585 & 14.599 & 0.244 & 0.025 & 0.022 & 3.457 & 6.994 & 3.857 & 668\\
V87 & 0.396488 & 14.592 & 14.605 & 0.232 & 0.017 & 0.015 & 3.446 & 7.400 & 4.416 & 727\\
V95 & 0.405067 & 14.559 & 14.568 & 0.211 & 0.018 & 0.017 & 3.834 & 7.380 & 4.371 & 549\\
V98 & 0.280566 & 14.782 & 14.796 & 0.237 & 0.039 & 0.021 & 3.199 & 6.143 & 3.412 & 703\\
V103 & 0.328852 & 14.538 & 14.544 & 0.150 & 0.014 & 0.005 & 3.384 & 6.313 & 3.383 & 711\\
V105 & 0.335328 & 14.745 & 14.760 & 0.247 & 0.025 & 0.018 & 3.434 & 7.087 & 4.104 & 716\\
V117 & 0.421641 & 14.473 & 14.484 & 0.215 & 0.016 & 0.016 & 4.012 & 7.547 & 4.722 & 572\\
V119 & 0.305876 & 14.686 & 14.692 & 0.154 & 0.017 & 0.004 & 3.263 & 6.189 & 3.056 & 567\\
V121 & 0.304182 & 14.588 & 14.593 & 0.139 & 0.015 & 0.003 & 3.160 & 6.098 & 2.963 & 569\\
V124 & 0.331860 & 14.560 & 14.577 & 0.257 & 0.047 & 0.025 & 3.278 & 6.171 & 3.244 & 733\\
V126 & 0.341891 & 14.600 & 14.618 & 0.243 & 0.041 & 0.027 & 3.219 & 6.484 & 3.508 & 139\\
V137 & 0.334205 & 14.548 & 14.561 & 0.240 & 0.040 & 0.023 & 3.212 & 6.305 & 3.345 & 587\\
V153 & 0.386245 & 14.577 & 14.589 & 0.218 & 0.019 & 0.015 & 3.336 & 7.237 & 4.167 & 729\\
V155 & 0.413925 & 14.516 & 14.526 & 0.206 & 0.018 & 0.016 & 3.827 & 7.284 & 4.347 & 709\\
V158 & 0.367276 & 14.508 & 14.516 & 0.174 & 0.011 & 0.008 & 3.621 & 7.431 & 4.543 & 740\\
V163 & 0.313229 & 14.554 & 14.557 & 0.111 & 0.008 & 0.002 & 3.294 & 6.623 & 3.621 & 593\\
V168 & 0.321299 & 15.135 & 15.147 & 0.230 & 0.029 & 0.019 & 3.151 & 6.542 & 3.917 & 587\\
V169 & 0.319116 & 14.641 & 14.644 & 0.110 & 0.007 & 0.002 & 3.376 & 6.678 & 4.137 & 712\\
V184 & 0.303370 & 14.660 & 14.663 & 0.114 & 0.008 & 0.002 & 2.976 & 6.661 & 3.103 & 729\\
V264 & 0.321398 & 14.742 & 14.752 & 0.210 & 0.019 & 0.015 & 3.217 & 7.176 & 4.017 & 567\\
V266 & 0.352303 & 14.509 & 14.511 & 0.092 & 0.004 & 0.002 & 3.351 & 6.727 & 2.328 & 588\\
V267 & 0.315822 & 14.486 & 14.490 & 0.120 & 0.009 & 0.003 & 3.390 & 6.605 & 3.440 & 589\\
V270 & 0.312714 & 14.546 & 14.549 & 0.095 & 0.006 & 0.001 & 3.109 & 6.026 & 1.766 & 734\\
V272 & 0.311482 & 14.664 & 14.666 & 0.086 & 0.005 & 0.001 & 3.369 & 7.174 & 3.871 & 591\\
V273 & 0.367106 & 14.545 & 14.550 & 0.131 & 0.008 & 0.004 & 3.464 & 7.283 & 4.395 & 718\\
V274 & 0.311089 & 14.578 & 14.581 & 0.109 & 0.009 & 0.002 & 3.281 & 6.289 & 3.313 & 591\\
V285 & 0.329014 & 14.537 & 14.540 & 0.095 & 0.004 & 0.001 & 3.443 & 7.289 & 4.205 & 592\\
V289 & 0.308090 & 14.624 & 14.628 & 0.127 & 0.011 & 0.004 & 3.124 & 6.739 & 3.613 & 755\\
NV341 & 0.306136 & 14.454 & 14.458 & 0.128 & 0.013 & 0.004 & 3.065 & 6.110 & 3.107 & 721\\
NV343 & 0.310211 & 14.563 & 14.567 & 0.132 & 0.013 & 0.004 & 3.276 & 6.354 & 3.483 & 695\\
NV344 & 0.313764 & 14.605 & 14.607 & 0.041 & 0.001 & 0.000 & 3.462 & 4.819 & 1.226 & 591\\
NV346 & 0.327623 & 14.497 & 14.506 & 0.190 & 0.027 & 0.015 & 3.225 & 5.930 & 2.949 & 719\\
NV347 & 0.328849 & 14.479 & 14.492 & 0.225 & 0.041 & 0.019 & 3.107 & 6.156 & 3.309 & 585\\
NV350 & 0.379108 & 14.485 & 14.496 & 0.212 & 0.019 & 0.018 & 3.482 & 7.226 & 4.097 & 727\\
NV354 & 0.419934 & 14.537 & 14.547 & 0.208 & 0.014 & 0.014 & 4.037 & 7.649 & 4.607 & 589\\
\hline
\end{tabular}
\end{minipage}
\end{table*}

\begin{table*}
 \centering
 \begin{minipage}{200mm}
  \caption{Formulae for $M_V$ of RR$c$ stars with their ${\cal D}$
parameters and the resulting distance moduli to $\omega$ Cen with
its standard deviations\label{t:4}}
  \begin{tabular}{rlccc}
\hline \hline No. & Equation & ${\cal D}$ & $\mu_{\omega Cen}$ &
$\sigma$ \\ \hline \hline 1 & $M_V = -0.259 - 1.338 \log P + 0.726
A_1$ & 0.0998 & 14.0883 & 0.0140 \\ 2 & $M_V = -0.318 - 1.330 \log P
+ 4.620 A_3 + 0.042 \phi_{21}$ & 0.0987 & 14.0875 & 0.0138 \\ 3 &
$M_V = -0.653 - 2.025 \log P - 6.981 A_2 + 14.639 A_3 + 0.048
\phi_{21}$ & 0.0954 & 14.0892 & 0.0131 \\ \hline \hline
\end{tabular}
\end{minipage}
\end{table*}

\subsubsection{RRc as standard candles in LMC}\label{s322}

The question whether the shape of the RRc light curve is or is not
correlated to the luminosity in principle does not exclude their
use as standard candles. More problems in this respect stem
from the large intrinsic scatter of their magnitudes, of order 0.1
mag. For LMC \.Zebru\'n et al. (2001) list OGLE observations of
450 RR$c$ stars from the fields SC2, SC3, SC4 and SC5. As these
observations were only a byproduct of a project designed for other
purposes, the number of $V$ filter observations, their exposure
length and distribution are not optimal for measuring such fine
effects as the third harmonic in the small amplitude, near
sinusoidal light curves of RRc. For this practical reason we could
not test Simon \& Clement (1993) formula (Eq. \ref{e:11}) as none
of the OGLE light curves yielded $\phi_{31}$ with the required
accuracy of 0.2 radians.

Application of our formula (Eq. \ref{e:12}) for these data was
straightforward, as finding the amplitude of the OGLE $V$ light curves
posed no difficulty at all. From the total sample of 450 RR$c$
stars we selected only 57 variables with sufficient amplitude
$A_1>0.1$ mag and errors not exceeding 0.010 mag. In Fig. 7 we
plot $M_V$ computed from Eq. (\ref{e:12}) against the observed
average magnitude. In this respect we were reassured, in a
perverse way, that the LMC RRc stars behave quite similarly to our
stars from $\omega$ Cen: they both reveal little correlation of
the predicted $M_V$ with its actual value. Such a situation calls
for a repeated analysis of the type performed by KW01 for as
large a sample of RR$c$ stars from different globular clusters as
possible, in order to check whether the formulae based on Fourier
coefficients have any prediction value for $M_V$. The
impression of similar properties is further confirmed in the
$P-A_1$ diagram (Fig. 8) plotted for RR$c$ stars from both
$\omega$ Cen (full circles) and the LMC (open circles). Both
samples occupy the same locations without showing any systematic
differences. In our opinion, this particular similarity justifies
use of the RRc stars as standard candles in these clusters.

\begin{figure}
 \vspace{7cm}
 \caption{The $M_V - <V>$ relation for the 57 OGLE RR$c$ variables
from the LMC. The solid line has a slope of unity.}
\includegraphics{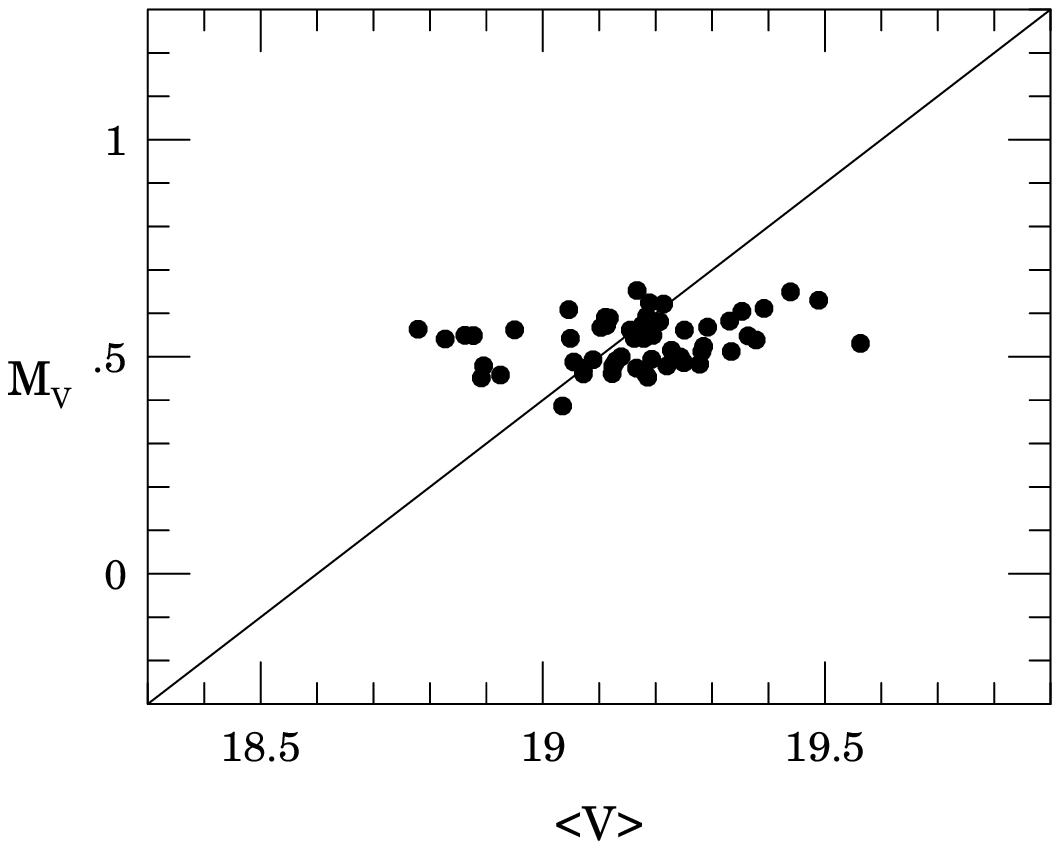}
\end{figure}

\begin{figure}
 \vspace{7cm}
 \caption{The $A_1 - P$ relation for the 57 OGLE RR$c$ variables
from the LMC (open circles) and 55 CASE RR$c$ variables from $\omega$ Cen.}
\includegraphics{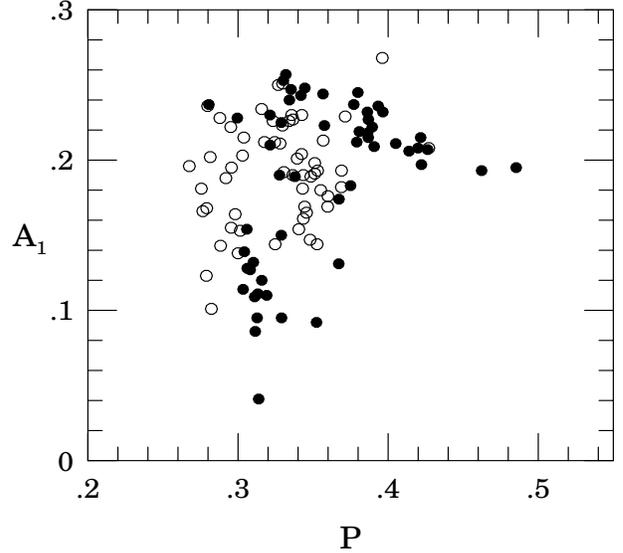}
\end{figure}

\section{Discussion}

The $V$-band photometry of stars in $\omega$ Cen obtained by
Kaluzny et al. (2003) contains 3 times as many observations and with
higher precision than available so far. To understand what could
be learned from such data we attempted here to redo from scratch
analysis of the RR Lyr stars in this cluster. The results combined
with the new metallicity determinations of Rey et al. (2000) and
distance modulus to the cluster derived by Kaluzny et al. (2002),
yield for RR Lyr stars a new $M_V - {\rm [Fe/H]}$ relation
independent of previous ones. For evolved stars, with $P>0.7$ days
the relation becomes nonlinear. Rejecting them, we obtain a nearly
linear relation with inclination consistent with previous results.
Applying this relation to available observations of RR Lyr in the
Large Magellanic Cloud we obtain its distance modulus
$\mu_{LMC}=18.56\pm0.14$ mag consistent with the "long" distance
scale.

Using the same $V$-band photometry we derive and calibrate the
formulae connecting the absolute magnitude of RR Lyr stars of
Bailey type $ab$ with the Fourier parameters describing their
light curves. This approach, closely following previous works,
proved only partially successful in that the residuals from fit
remain large compared to the statistical errors of observations
and in that the new formulae show little relation to the ones
derived from older data. Employing our best fit formula calibrated
with our distance modulus of $\omega$ Cen to the OGLE photometry
of 13 RR Lyr stars from the LMC enabled us to obtain its distance
modulus equal to $\mu_{LMC}=18.51\pm0.07_r\pm0.04_s$ mag.

The agreement with the value obtained from the $M_V - {\rm [Fe/H]}$
relation is very good, which on one hand is hardly surprising as both
data sets overlap, yet on the other hand demonstrates the formal
validity of our methods. Summarizing, we estimate the LMC distance
modulus equal to $\mu_{LMC}=18.52\pm0.06_r\pm0.04_s$ mag where the
systematic error is determined by the error in the distance modulus of
$\omega$ Cen (Kaluzny et al. 2002). The distance to the LMC is thus
equal to $50.6 \pm 1.6$ kpc.

Our results for the RR$c$ stars are less encouraging. Despite nearly
sinusoidal light curves, our photometry proved to be accurate enough for
analyzing the harmonics, however any fitted formulae employing Fourier
coefficients yielded $M_V$ only loosely correlated with $<V>$. On the
theoretical side this indicates dependence of $<V>$ on some additional
factor yet to be identified. On the practical side any estimated $M_V$
suffer from large random errors. A comparison with the RRc stars from
the LMC observed by the OGLE team (\.Zebru\'n et al. 2001) proved
difficult, because apart from the intrinsic errors of our formulae, the
poorer sampling of data, obtained for other purposes, prevented an
accurate evaluation of the harmonics.

An excellent review of the recent distance determinations to the LMC was
given by Walker (1999). Here we point out only the discrepancy between
so-called "short" and "long" distance scales. The former, based on
the statistical parallax and the red clump methods, gives the distance
modulus to the LMC as approximately 18.2-18.3 mag (Gould \& Popowski
1998, Udalski 2000a, 2000b). This is 0.2-0.3 mag smaller than the value
of about 18.50 mag resulting from the cepheid period-luminosity
relation and theoretical models of the horizontal branch or globular
clusters main sequence fitting.

However, recent improvements of the red clump method based on the
infrared photometry seem to indicate that this method also provides the
distance modulus of the LMC at around 18.50 mag. For example, three
papers based on the $K$-band photometry were recently published giving
the distance modulus to the LMC as equal to $18.49\pm0.04$,
$18.54\pm0.10$ and $18.501\pm0.008_r\pm0.045_s$ mag, respectively (Alves
et. al 2002, Sarajedini et al. 2002, Pietrzy\'nski \& Gieren 2002).

Also, the newest determination of the cepheid period-luminosity relation,
based on $\sim$600 stars, indicates that the distance to the LMC is
around 18.50 mag (Sebo et al. 2002).

Our estimate is in excellent agreement with these recent results proving
that our calibration of the absolute magnitudes of RR Lyr stars produces
valuable results.

\section*{Acknowledgments}

We would like to thank Igor Soszy\'nski from the OGLE project for his
valuable suggestions concerning the OGLE Catalogue of Variable Stars in
the LMC and Grzegorz Stachowski for reading the manuscript. This work
was supported by the KBN grant number 2~P03D~024~22 to A. Olech and
5~P03D~004~21 to J. Kaluzny.


\begin{thebibliography}{99}
\bibitem{b1} Alves D.R., Rejkuba M., Minniti D., Cook K.M., 2002, ApJ,
573, L51
\bibitem{b1} Butler D., Dickens R.J., Epps E.A., 1978, ApJ, 225, 148
\bibitem{b1} Caloi V., D'Antona F., Mazzitelli I., 1997, A\&A, 320, 823
\bibitem{b1} Caputo F., 1997, MNRAS, 284, 994
\bibitem{b1} Caputo F., Castellani V., Marconi M., Ripepi V., 2000,
MNRAS, 316, 819
\bibitem{b1} Chaboyer B., 1999, "Globular cluster distance
determinations" in "Post-Hipparcos cosmic candles", Kluwer Academic
Publishers, eds. A. Heck, F. Caputo, p. 111
\bibitem{b1} Clementini G., Gratton R., Bragaglia A., Carreta E.,
Di Fabrizio L., Maio M., 2003, AJ, 125, 1309
\bibitem{b1} Dutra C.M., Bica E., Clari\'a J.J., Piatti A.E., Ahumada
A.V., 2001, A\&A, 371, 895
\bibitem{b1} Fearnley J., Carney B.W., Skillen I., Cacciari C., Janes
K., 1998, MNRAS, 293, L61
\bibitem{b1} Fitzpatrick E.L., Ribas I., Guinan E.F., DeWarf, L.E.,
Maloney F.P., Massa D., 2002, ApJ, 564, 260
\bibitem{b1} Fusi Pecci F., Buonanno R., Cacciari C., Corsi C.E., Djorgovski
S.G., Federici L., Ferraro F.R., Permeggiani G., Rich R.M., 1996, AJ,
112, 1461
\bibitem{b1} Gould A., Popowski P., 1998, ApJ, 508, 844
\bibitem{b2} Gratton R.G., Fusi Pecci F., Carretta E., Clementini G.,
Corsi C.E., Lattanzi M., 1997, ApJ, 491, 749
\bibitem{b1} Groenewegen M.A.T., Salaris M., 2001, A\&A, 366, 752
\bibitem{b1} Harris W.E., 1996, AJ, 112, 1487
\bibitem{b1} Kaluzny J., Kubiak M., Szyma\'nski M., Udalski A., Krzemi\'nski
W., Mateo M., 1997, A\&AS, 122, 471
\bibitem{b2} Kaluzny J., Thompson I.B., Krzemi\'nski W., Olech A., Pych
W., Mochejska B., 2002, ASP Conf. Ser. 265, 155
\bibitem{b2} Kaluzny J., Olech A., Thompson I.B., Krzemi\'nski W., Pych W.,
Schwarzenberg-Czerny A., 2003, A\&A, submitted A.
\bibitem{b3} Kov\'acs G., Jurcsik J., 1996, ApJ, 466, L17
\bibitem{b3} Kov\'acs G., Kanbur S.M., 1998, MNRAS, 295, 834
\bibitem{b3} Kov\'acs G., Walker A.R., 2001, A\&A, 371, 679, KW01
\bibitem{b3} Lee Y.,-W., 1990, ApJ, 363, 159
\bibitem{b3} Lee Y.,-W., 1991, ApJ, 373, l43
\bibitem{b3} Ligeza, P.; Schwarzenberg-Czerny, A.
   2000, International Astronomical Union. Symposium no. 201, p71.
\bibitem{b3} Pietrzy\'nski G., Gieren W., 2002, AJ, 124, 2633
\bibitem{b3} Rey S.-C., Lee Y.-W., Joo J.-M., Walker A., Baird S., 2000,
AJ, 119, 1824
\bibitem{b3} Sandage A.R., 1981a, ApJ, 244, L23
\bibitem{b4} Sandage A.R., 1981b, ApJ, 248, 161
\bibitem{b4} Sarajedini A., Grocholski A.J., Levine J., Lada E., 2002,
AJ, 124, 2625
\bibitem{b4} Schlegel. D.J., Finkbeiner, D.P., Davis, M., 1998, ApJ,
500, 525
\bibitem{b4} Schwarzenberg-Czerny, A. 1996, ApJ, 460, L107.
\bibitem{b4} Schwarzenberg-Czerny, A.  \& Kaluzny, J.
     1998, MNRAS, 300, 251.
\bibitem{b4} Sebo K.M., et al., 2002, ApJ Suppl. Ser., 142, 71
\bibitem{b4} Simon N.R., 1989, ApJ, 343, L17
\bibitem{b4} Simon N.R., Teays T.J., 1982, ApJ, 261, 586
\bibitem{b4} Simon N.R., Clement C.M., 1993, ApJ, 410, 526
\bibitem{b3} Thompson I.B., Kaluzny J., Pych W., Burley G., Krzeminski
W., Paczy\'nski B., Persson S.E., Preston G.W., 2001, AJ, 121, 3089
\bibitem{b5} Udalski A., 2000a, Acta Astronomica, 50, 279
\bibitem{b5} Udalski A., 2000b, ApJ, 531, L25
\bibitem{b5} van Leeuwen F., Le Poole R.S., Reijns R.A., Freeman K.C., 
de Zeeuw P.T., 2000, A\&A, 360, 472 
\bibitem{b5} Walker A.R., 1992, ApJ, 390, L81
\bibitem{b5} Walker A.R., 1999, "Distance to the Magellanic Clouds" in
"Post-Hipparcos cosmic candles", Kluwer Academic Publishers, eds. A.
Heck, F. Caputo, p. 100
\bibitem{b6} Zinn R., West M.J., 1984, ApJS, 55, 45
\bibitem{b6} \.Zebru\'n K., Soszy\'nski I., Wo\'zniak P.R., Udalski A.,
Kubiak M., Szyma\'nski M., Pietrzy\'nski G., Szewczyk O., Wyrzykowski
{\L}., 2001, Acta Astronomica, 51, 317

\end{thebibliography}
\end{document}